\documentclass[longbibliography,showpacs,floatfix,nofootinbib,superscriptaddress,twocolumn,aps,prx]{revtex4-2}

\usepackage{graphicx,color}
\usepackage{amsfonts}
\usepackage[figuresright]{rotating}  
\usepackage{amssymb}
\usepackage{bm}
\usepackage{amsmath}
\usepackage{mathtools}
\usepackage{psfrag}
\usepackage{multirow}
\usepackage{tabularx}
\usepackage{textcomp}
\usepackage{units}
\usepackage{lipsum}
\usepackage{wasysym} 
\usepackage{graphicx}
\usepackage{physics}
\usepackage{soul}
\usepackage{times}
\usepackage{enumitem}
\usepackage{floatrow}
\usepackage[dvipsnames]{xcolor}
\usepackage[caption=false]{subfig}
\usepackage{tikz}
\usepackage{appendix}
\usepackage{dsfont}
\usepackage{mathrsfs}
\usepackage[colorlinks=true,citecolor=blue,linkcolor=magenta,hypertexnames=false]{hyperref}

\renewcommand{\selectlanguage}[1]{}

\def \kuohao#1{\left(#1\right)}
\def \fkuohao#1{\left[#1\right]}
\def \hkuohao#1{\left\{#1\right\}}
\def \abs#1{\left|#1\right|}
\def \diag#1{\mathrm{diag}\kuohao{#1}}
\newcommand{\neiji}[2]{\langle#1|#2\rangle}
\newcommand{\qiwang}[1]{\langle#1\rangle}

\begin{document}

\title{Fermionic Partial Transpose in the Overlap Matrix Framework for Entanglement Negativity}

\author{Jun Qi Fang}
\affiliation{Key Laboratory of Artificial Structures and Quantum Control (Ministry of Education), School of Physics and Astronomy, Shanghai Jiao Tong University, Shanghai 200240, China}
\author{Xiao Yan Xu}
\email{Contact author: xiaoyanxu@sjtu.edu.cn}
\affiliation{Key Laboratory of Artificial Structures and Quantum Control (Ministry of Education), School of Physics and Astronomy, Shanghai Jiao Tong University, Shanghai 200240, China}
\affiliation{Hefei National Laboratory, Hefei 230088, China}

\begin{abstract}
Over the past two decades, the overlap matrix approach has been developed to compute quantum entanglement in free-fermion systems, particularly to calculate entanglement entropy and entanglement negativity. This method involves the use of partial trace and partial transpose operations within the overlap matrix framework. However, in previous studies, only the conventional partial transpose in fermionic systems has been considered, which does not account for fermionic anticommutation relations. Although the concept of a fermionic partial transpose was introduced by Shapourian \textit{et al}. [\href{https://doi.org/10.1103/PhysRevB.95.165101}{Phys. Rev. B \textbf{95}, 165101 (2017)}], it has not yet been systematically incorporated into the overlap matrix framework. In this paper, we introduce the fermionic partial transpose into the overlap matrix approach, provide a systematic analysis of the validity of partial trace and partial transpose operations, and derive an explicit formula for calculating entanglement negativity in bipartite systems. Additionally, we numerically compute the logarithmic negativity of two lattice models to verify the Gioev-Klich-Widom scaling law. For tripartite geometries, we uncover limitations of the overlap matrix method and demonstrate that the previously reported logarithmic negativity result for a homogeneous one-dimensional chain in a disjoint interval geometry exceeds its theoretical upper bound.
\end{abstract}
\maketitle

\section{Introduction}
Research on quantum entanglement in recent decades has offered an additional perspective on the analysis of many-body states in condensed matter physics \cite{Kitaev_critical,entanglement_many-body_system,entanglement_condensed_matter_system,EE_CFT}. Unlike classical correlations, quantum entanglement reveals the intrinsic connections between subsystems of a quantum state. This unique property has made quantum entanglement a critical framework for understanding collective behaviors, quantum phase transitions, topological order, and quantum criticality in condensed matter systems \cite{colloquium_of_area_law,concurrence_scaling_in_phase_transition,Kitaev_topo_entropy,topo_Wen,singularity_negativity_finite_temp}. Among the various measures of entanglement, the entanglement entropy has proven to be particularly effective in characterizing bipartite entanglement in the ground states of many-body systems \cite{EE_extended_quantum,topo_order_EE,EE_CFT}.

However, for mixed states, such as tripartite systems and thermal states, entanglement entropy alone cannot fully characterize entanglement, as it also includes contributions from classical correlations. To address this limitation, several measures have been proposed to quantify entanglement in mixed states \cite{quantum_entanglement_RMP}. Among these, entanglement negativity \cite{negativity_original,log_negativity_original,comparison_formaton_negativity}, based on the partial transpose operation, is widely used. One of the key features of this measure is that, for separable states, the eigenvalues of the density matrix remain non-negative after applying the partial transpose operation \cite{ppt_operation,condition_ppt}, making it a powerful diagnostic tool for identifying entanglement.

Compared with other mixed state entanglement measures, entanglement negativity has the advantage of requiring only straightforward linear algebra computations. In this paper, we focus on \textit{logarithmic negativity}, which is defined as follows. Given the density matrix of a mixed state, such as a reduced density matrix obtained by tracing out region $B$ while retaining $\rho_{A_{1}\cup A_{2}}$, the logarithmic negativity is given by
\begin{equation}
    \mathcal{E}=\ln\text{Tr}\abs{\rho_{A_{1}\cup A_{2}}^{T_{A_{2}}}},
\end{equation}
where $\rho_{A_{1}\cup A_{2}}^{T_{A_{2}}}$ denotes the partial transpose operation on region $A_2$ over the reduced density matrix $\rho_{A_{1}\cup A_{2}}$, and the trace norm operation $\text{Tr}\abs{O}$  represents the sum of the square roots of the eigenvalues of $O^{\dag} O$. If $O$ is Hermitian, the trace norm simplifies to the sum of the absolute eigenvalues of $O$. Logarithmic negativity has been extensively applied to investigate entanglement in various many-body systems, including one-dimensional harmonic oscillators \cite{Audenaert2002pra,Eisler2014,Ferraro2008prl,Nobili2016}, spin systems \cite{Bayat2010prl,Bayat2010prb,Bayat2012prl,Vidal2010pra,Wichterich2009pra,Hastings2010prl,Santos2011pra,Wu2020prl}, topologically ordered phases \cite{Castelnovo2013pra,Vidal2013pra,Grover2020prl,singularity_negativity_finite_temp}, and conformal field theory (CFT) \cite{Calabrese2013,Calabrese2012prl,boson_EN_CFT,finite_temp_EN}. 

The negativity depends on the definition of partial transpose. In the literature,
three distinct forms of partial transpose have been studied: bosonic partial transpose (bPT), as discussed in Ref. \cite{Eisler2015,Eisert2018prb}; untwisted partial transpose (uPT), introduced in Ref. \cite{Shapourian2017prb,Shapourian2018prb,Shapourian2019pra,fohong2023arxiv}; and twisted partial transpose (tPT), presented in Ref. \cite{Shapourian2019sp}.
In fermionic systems, the definition of partial transpose must respect the fermionic anticommuting relations. Both uPT and tPT satisfy this requirement, whereas bPT does not.  Compared with bPT, uPT and tPT have advantageous properties. For example, it has been shown that a fermionic Gaussian state remains Gaussian after applying uPT or tPT~\cite{Shapourian2017prb,Shapourian2018prb,Shapourian2019pra,Shapourian2019sp}, making them convenient for simulating R\'enyi negativity using the quantum Monte Carlo (QMC) algorithm~\cite{fohong2023arxiv,fohong2025}.  Interestingly, the logarithmic negativity calculated using tPT coincides with that of uPT. Therefore, in this work, we focus solely on uPT and omit further discussion of tPT.

However, in practice, the calculation of logarithmic negativity is very challenging, even for free fermion systems. For example, the Green function approach  \cite{Eisler2015,Eisert2018prb,Shapourian2017prb,Shapourian2019sp,liu2022entanglementnegativityversusmutual,footnote} for free-fermion systems is very powerful, but it is still difficult to  extract analytical properties of logarithmic negativity.
In contrast, the overlap matrix approach has emerged as a more effective tool for computing entanglement entropy and entanglement negativity of pure states in free-fermion systems  \cite{overlap_for_SE,overlap_for_Sn,overlap_for_EN,overlap_for_non-Hermite,overlap_supp1,kruchkov2024arxiv}. However, in previous applications of the overlap matrix approach, only bPT was considered.
In this paper, we incorporate uPT into the overlap matrix approach. We provide a systematic proof of the validity of the uPT operation, as well as a numerical and analytical confirmation of the partial trace operation in the bipartite geometry of free-fermion systems, and propose analytical results consistent with the CFT method.
Additionally, in the tripartite case, we demonstrate that the previous overlap matrix approach for computing logarithmic negativity exceeds its theoretical upper bound. 

This paper is organized as follows. In Sec. \ref{sec:bipartite entanglement review}, we introduce the fermionic partial transpose in the overlap matrix approach for entanglement negativity calculation in free fermions systems. We then present a rigorous proof that establishes the validity of this framework for bipartite pure states. Specifically, we demonstrate that the partially transposed density matrix from the overlap matrix method is related to its Fock space counterpart through a similarity transformation. In Sec. \ref{sec:bipartite entanglement proof and numerical result}, we analyze two illustrative examples, where our numerical results show excellent agreement with the Gioev-Klich-Widom scaling law. In Sec. \ref{sec:tripartite entanglement}, we study the logarithmic negativity of free fermions in the tripartite case, and demonstrate that in mixed states, the equivalence of partial transpose in the overlap matrix approach and in Fock space no longer holds. Consequently, we show that previously reported results for tripartite geometries violate their theoretical upper bound. Finally, in Sec. \ref{sec:conclusion}, we summarize our findings.

\section{Overlap matrix approach in free-fermion systems}\label{sec:bipartite entanglement review}
We begin with a brief review of the original overlap matrix approach and introduce the fermionic partial transpose operation for entanglement negativity calculation. Consider a general Hermitian Hamiltonian of a free-fermion system
\begin{equation}\label{general free fermion}
    H =\sum_{i,j}h_{i,j}c_i^\dagger c_j 
      =\sum_{\alpha=1}^N\epsilon_\alpha f_\alpha^\dagger f_\alpha,
\end{equation}
where the Hermitian matrix $h$ can be diagonalized by a unitary matrix $U$, such that $f_{\alpha}^\dagger=\kuohao{\textbf{c}^{\dag}U}_{\alpha}=\sum_{i=1}^{N}c_{i}^{\dag}U_{i,\alpha}$ is the fermionic operator in the diagonal basis with energy $\epsilon_{\alpha}$, satisfying the anticommutation relation $\hkuohao{f_{\alpha},f_{\beta}^{\dag}}=\delta_{\alpha,\beta}$. Here, $N$ denotes the number of lattice sites.
An $M$-particle ground state can be written as 
\begin{equation}
    \ket{\Psi}=\prod_{\alpha=1}^{M}f_{\alpha}^{\dag}\ket{0},
\end{equation}
where $M$ lowest energy single particle states are occupied. 

\subsection{Methodology}
We consider a bipartite system in which the lattice sites are divided into two regions, labeled by $A$ and $B$. A projection operator $\mathcal{P}_{A\kuohao{B}}$ can be defined to only keep the particles in $A\kuohao{B}$ for a given state. Specifically, the projection of the $f^{\dag}_{\alpha}$ operator satisfies $\mathcal{P}_Af_\alpha^\dagger=\sum_{i\in A}c_i^\dagger U_{i,\alpha}$ and $\mathcal{P}_Bf_\alpha^\dagger=\sum_{i\in B}c_i^\dagger U_{i,\alpha}$. As expected, the sum of the projections satisfies $\mathcal{P}_{A}f_{\alpha}^{\dag}+\mathcal{P}_{B}f_{\alpha}^{\dag}=f_{\alpha}^{\dag}$, ensuring completeness. An overlap matrix is then introduced
\begin{equation}
\begin{aligned}
    \fkuohao{M_{A}}_{\alpha,\beta}&=\neiji{\mathcal{P}_{A}u_{\alpha}}{\mathcal{P}_{A}u_{\beta}}\\
    & =\sum_{i\in A}U_{i\alpha}^*U_{i\beta},\quad 1\leq \alpha,\beta\leq M
\end{aligned}
\end{equation}
where $\ket{u_{\alpha}}=f_{\alpha}^{\dag}\ket{0}$ is a single-particle state. It can be easily verified that $M_{A\kuohao{B}}$ is Hermitian and $M_{A}+M_{B}=\mathds{1}$. Consequently, the two overlap matrices $M_{A}$ and $M_{B}$ can be simultaneously diagonalized as $\mathcal{U}^{\dag}M_{A}\mathcal{U}=\diag{\hkuohao{P_{\gamma}}_{\gamma=1}^{M}}$ and $\mathcal{U}^{\dag}M_{B}\mathcal{U}=\diag{\hkuohao{1-P_{\gamma}}_{\gamma=1}^{M}}$. The eigenvalues $P_{\gamma}$ are restricted to the range $\fkuohao{0,1}$ (see Appendix \ref{app_a} for a proof). We can use $\mathcal{U}$ to define a new basis 
\begin{equation}
d_{\gamma}^{\dagger} = \sum_{\alpha=1}^{M}f_{\alpha}^{\dagger}\mathcal{U}_{\alpha,\gamma}.
\end{equation}
In this new basis, the former $M$-particle ground state becomes
\begin{equation}
\label{eq:psi_in_om_basis}
\ket{\Psi}=\exp\kuohao{\text{i}\theta}\prod_{\gamma=1}^Md_\gamma^\dagger\ket{0},
\end{equation}
where the phase factor $\exp\kuohao{\text{i}\theta}=\det\kuohao{\mathcal{U}^{\dag}}$. Another interesting property of this new basis is that the region $A$ and $B$ part operators separate
\begin{equation}
d_{\gamma}^{\dagger}\equiv\sqrt{P_{\gamma}}d_{A_{\gamma}}^{\dagger}+\sqrt{1-P_{\gamma}}d_{B_{\gamma}}^{\dagger}
\end{equation}
where
\begin{equation}
    d_{A\gamma}^{\dag}=\frac{\sum_{\alpha}\mathcal{U}_{\alpha \gamma}\mathcal{P}_{A}f_{\alpha}^{\dag}}{\sqrt{P_{\gamma}}},\ d_{B\gamma}^{\dag}=\frac{\sum_{\alpha}\mathcal{U}_{\alpha \gamma}\mathcal{P}_{B}f_{\alpha}^{\dag}}{\sqrt{1-P_{\gamma}}},
\end{equation}
and they preserve the anticommutation relations $\hkuohao{d_{A\kuohao{B}\gamma},d_{A\kuohao{B}\gamma'}^{\dag}}=\delta_{\gamma\gamma'}$ and $\hkuohao{d_{A\kuohao{B}\gamma}^{\kuohao{\dag}},d_{B\kuohao{A}\gamma'}^{\kuohao{\dag}}}=0$.

Since each term in the product of Eq.~\eqref{eq:psi_in_om_basis} is independent, it follows that the system can be described as a tensor product, and consequently, the density matrix also factorizes accordingly
\begin{equation}
    \rho=\bigotimes_{\gamma=1}^{M}\begin{bmatrix}
        1-P_{\gamma} & \sqrt{P_{\gamma}\kuohao{1-P_{\gamma}}}\\
        \sqrt{P_{\gamma}\kuohao{1-P_{\gamma}}} & P_{\gamma}
    \end{bmatrix}\equiv \bigotimes_{\gamma=1}^{M}\rho_{\gamma},
\end{equation}
where the basis is ordered as $\hkuohao{\ket{0_{A\gamma}1_{B\gamma}},\ket{1_{A\gamma}0_{B\gamma}}}$ and the creation operator acts as $d_{A\kuohao{B}\gamma}^{\dag}\ket{0}=\ket{1_{A\kuohao{B}\gamma}}$. Notably, the original density matrix, which initially has dimensions $2^{N}\times 2^{N}$, is effectively reduced to $2^{M}\times 2^{M}$. This reduction occurs because other rays in the Hilbert space do not contribute to $\ket{\Psi}$, and all other matrix elements in $\rho$ are strictly zero. 

The partial trace in $d^{\dag}$ representation yields
\begin{equation}\label{pt in d+}
\begin{aligned}
    \rho_{A}&=\bigotimes_{\gamma=1}^{M}\begin{bmatrix}
        1-P_{\gamma} & \\
        & P_{\gamma}
    \end{bmatrix}\\
    &=\bigotimes_{\gamma=1}^{M}\rho_{A\gamma},
\end{aligned}
\end{equation}
where the basis is ordered as $\hkuohao{\ket{0_{A\gamma}},\ket{1_{A\gamma}}}$. Before performing the partial transpose, it is useful to outline some key properties of the uPT, as discussed in Ref. \cite{Shapourian2019pra}. The definition in Fock space is given by
\begin{equation}
\begin{aligned}
    &\hspace{1em}(|\{n_j\}_{A},\{n_j\}_{B}\rangle\langle\{\bar n_j\}_{A},\{\bar n_j\}_{B}|)^{T_B^{f}}\\
    &=(-1)^{\phi(\{n_j\},\{\bar n_j\})}|\{n_j\}_{A},\{\bar n_j\}_{B}\rangle\langle\{\bar n_j\}_{A},\{n_j\}_{B}|,
\end{aligned}
\end{equation}
where the phase factor
\begin{equation}
    \phi(\{n_j\},\{\bar{n}_j\})=\frac{[(\tau_B+\bar{\tau}_B)\bmod2]}{2}+(\tau_A+\bar{\tau}_A)(\tau_B+\bar{\tau}_B),
\end{equation}
where $\tau_{A\kuohao{B}}=\sum_{j\in A\kuohao{B}}n_{j}$ and $\bar{\tau}_{A\kuohao{B}}=\sum_{j\in A\kuohao{B}}\bar{n}_{j}$. This uPT: (1) preserves the tensor product structure of fermionic density matrices, whereas the bPT does not
\begin{equation}
    \kuohao{\rho_{AB}\otimes\rho'_{AB}}^{T_{A}^{f}}=\rho_{AB}^{T_{A}^{f}}\otimes\kuohao{\rho'_{AB}}^{T_{A}^{f}};
\end{equation}
(2) leaves the logarithmic negativity invariant under a unitary transformation
\begin{equation}\label{equivalence after unitary transformation}
    \mathcal{E}\kuohao{\rho}=\mathcal{E}\fkuohao{\kuohao{U_{A}\otimes U_{B}}\rho\kuohao{U_{A}^{\dag}\otimes U_{B}^{\dag}}};
\end{equation}
(3) satisfies the additivity property
\begin{equation}\label{additivity}
    \mathcal{E}\kuohao{\rho_{AB}\otimes\rho_{AB}^\prime}=\mathcal{E}\kuohao{\rho_{AB}}+\mathcal{E}\kuohao{\rho'_{AB}}.
\end{equation}
Utilizing these three essential properties, we demonstrate that the bPT for the density matrix in Ref. \cite{overlap_for_EN} is not appropriate, as it violates property (1) above if bPT is applied to fermionic systems~\cite{Shapourian2019pra}. To address this issue, we introduce the uPT to evaluate entanglement negativity. The uPT density matrix can be expressed as
\begin{equation}
    \rho^{T_{B}^f} =\bigotimes_{\gamma=1}^{M}\rho_{\gamma}^{T_{B}^{f}},
\end{equation}
with
\begin{equation}\label{upt in d+}
    \rho_{\gamma}^{T_{B}^{f}}=\begin{bmatrix}
        1-P_\gamma & & & \\
        & P_\gamma & &\\
        & & & -\text{i}\sqrt{P_\gamma(1-P_\gamma)}\\
        & & -\text{i}\sqrt{P_\gamma(1-P_\gamma)} & 
    \end{bmatrix}.
\end{equation}
where the basis is ordered as $\hkuohao{\ket{0_{Ai}1_{Bi}},\ket{1_{Ai}0_{Bi}},\ket{0_{Ai}0_{Bi}},\ket{1_{Ai}1_{Bi}}}$. The correctness of Eqs. (\ref{pt in d+}) and (\ref{upt in d+}) will be discussed in detail later. 

Due to the tensor product structure of the density matrix, both the entanglement entropy and entanglement negativity can be computed efficiently. As an example, consider the logarithmic negativity. The eigenvalues of the Kronecker product of two matrices, $A\otimes B$, are given by all possible products $\lambda_{\mu}\mu_{\nu}$, where $\lambda_{\mu}$ and $\mu_{\nu}$ are the eigenvalues of $A$ and $B$, respectively. For the non-Hermitian $\rho^{T_{B}^{f}}$, we need to compute the eigenvalues of $\sqrt{\kuohao{\rho^{T_{B}^{f}}}^{\dag}\rho^{T_{B}^{f}}}$, which are given by $\hkuohao{\varXi_{\gamma,\alpha}}=\hkuohao{1-P_{\gamma},P_{\gamma},\sqrt{P_{\gamma}\kuohao{1-P_{\gamma}}},\sqrt{P_{\gamma}\kuohao{1-P_{\gamma}}}}$. Thus, the logarithmic negativity can be formulated as
\begin{equation}\label{bipartite log negativity formula}
\begin{aligned}
    \mathcal{E}&=\ln\mathrm{Tr}|\rho^{T_{B}^{f}}|=\ln\prod_{\gamma}\sum_{\alpha}\Xi_{\gamma,\alpha}\\
    &=\sum_{\gamma}\ln\fkuohao{1+2\sqrt{P_{\gamma}(1-P_{\gamma})}}.
\end{aligned}
\end{equation}

Importantly, it is shown that this formula is identical to that of Ref.~\cite{overlap_for_EN}, even though they treated the fermionic system as a bosonic system. In contrast, our approach uses the appropriate uPT for fermionic systems, ensuring consistency with the underlying fermionic structure.
The formula for Von Neumann and Rényi entropy, consistent with Ref. \cite{overlap_for_SE,overlap_for_Sn}, can be derived similarly, as can the Rényi negativity.

\begin{figure}[!t]
    \centering
    \includegraphics[width=\columnwidth]{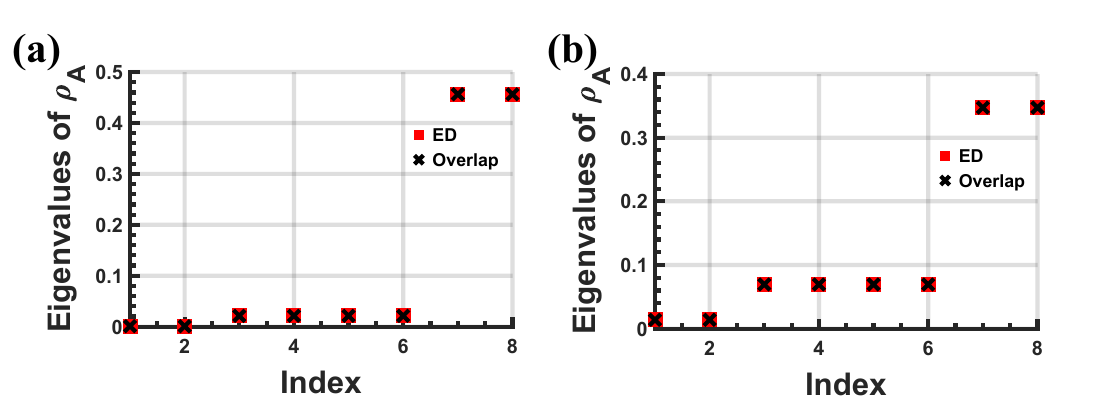}
    \caption{Comparison of the eigenvalues of $\rho_{A}$, after reordering and removing the zero components, between the partial trace in the overlap matrix approach and in Fock space. We consider an $N=12$ and $\mu/t=1$ homogeneous one-dimensional chain in two bipartite geometries: (a) an adjacent interval case where the system is evenly divided into regions $A$ and $B$; (b) a disjoint interval case where the system is evenly partitioned into $A_{1}$, $B_{1}$, $A_{2}$, and $B_{2}$ sequentially \cite{data_available}. }
    \label{fig:comparison between overlap matrix mapping and Fock space}
\end{figure}

\subsection{A general proof}
Incredibly when we perform both the partial trace and partial transpose in the new basis, the outcome appears to be identical to that observed in the Fock basis. A rigorous proof is therefore required to validate this equivalence.

\textit{Partial trace. }The reduced density matrix obtained by analytically performing the partial trace in the Fock basis  is compared with the reduced density matrix obtained by the overlap matrix method. The comparison reveals that these two matrices are identical. A detailed analytical derivation confirming this result is provided in Appendix \ref{app_d}. We also verify this identical relationship numerically by comparing the eigenvalues of Eq. (\ref{pt in d+}) with exact diagonalization results. We consider two types of bipartite geometries: adjacent intervals and disjoint intervals. As shown in Fig. \ref{fig:comparison between overlap matrix mapping and Fock space}, both geometries yield identical eigenvalues for the partial trace in different representations. 
Our analytical proof and numerical verifications justify that the entanglement spectrum \cite{Calabrese2008pra,Assaad2014prb,Yepeng2015prb,Hermanns2014jsm} can be reliably obtained using the overlap matrix approach.

The key reason behind the validity of the partial trace in the overlap matrix approach lies in its semilocal nature, meaning that it remains associated with either region $A$ or $B$. Consequently, tracing out region $A$ or $B$ does not lead to the loss of information about the remaining region, which is different from the Jordan-Wigner transformation \cite{partial_trace_in_JW1,partial_trace_in_JW2,partial_trace_in_JW3}. 

\textit{Partial transpose. }For the Hamiltonian in Eq. (\ref{general free fermion}), given that the density matrix of the ground state has a tensor product structure, and using Eq. (\ref{additivity}), the logarithmic negativity is
\begin{equation}
    \mathcal{E}\kuohao{\bigotimes_{\gamma}\rho_{\gamma}^{T_{B}^{f}}}=\sum_{\gamma}\mathcal{E}\kuohao{\rho_{\gamma}^{T_{B}^{f}}}.
\end{equation}
Hence, one can analyze each $\rho_{\gamma}^{T_{B}^{f}}$ individually. We prove that the partially transposed density matrix $\rho_{\gamma}$ in the overlap matrix approach is related to its counterpart in Fock space through a unitary transformation. We expand the partially transposed single-mode density matrix in the overlap matrix approach ${\rho}_{\gamma}^{T_{B}^{f}}$ into Fock space
\begin{equation}\label{rho_i^T in overlap matrix mapping}
\begin{aligned}
    {\rho}_{\gamma}^{T_{B}^{f}}&=\sum_{i\in A}\sum_{i'\in A}\kuohao{U\mathcal{U}}_{i\gamma}\kuohao{\mathcal{U}^{\dag}U^{\dag}}_{\gamma i'}c_{i}^{\dag}\ket{0}\bra{0}c_{i'}\\
    &\hspace{1em}+\sum_{i\in B}\sum_{i'\in B}\kuohao{U\mathcal{U}}_{i\gamma}\kuohao{\mathcal{U}^{\dag}U^{\dag}}_{\gamma i'}c_{i}^{\dag}\ket{0}\bra{0}c_{i'}\\
    &\hspace{1em}{-\text{i}}\sum_{i\in A}\sum_{i'\in B}\kuohao{\mathcal{U}^{\dag}U^{\dag}}_{\gamma i}\kuohao{\mathcal{U}^{\dag}U^{\dag}}_{\gamma i'}\ket{0}\bra{0}c_{i'}c_{i}\\
    &\hspace{1em}{-\text{i}}\sum_{i\in A}\sum_{i'\in B}\kuohao{U\mathcal{U}}_{i\gamma}\kuohao{U\mathcal{U}}_{i'\gamma}c_{i}^{\dag}c_{i'}^{\dag}\ket{0}\bra{0}.
\end{aligned}
\end{equation}
Comparing this with the density matrix partially transposed in Fock space, denoted as $\Tilde{\rho}_{\gamma}^{T_{B}^{f}}$
\begin{equation}\label{rho_i^T in Fock space}
\begin{aligned}
    \Tilde{\rho}_{\gamma}^{T_{B}^{f}}&=\sum_{i\in A}\sum_{i'\in A}\kuohao{U\mathcal{U}}_{i\gamma}\kuohao{\mathcal{U}^{\dag}U^{\dag}}_{\gamma i'}c_{i}^{\dag}\ket{0}\bra{0}c_{i'}\\
    &\hspace{1em}+\sum_{i\in B}\sum_{i'\in B}\kuohao{U\mathcal{U}}_{i\gamma}\kuohao{\mathcal{U}^{\dag}U^{\dag}}_{\gamma i'}c_{i}^{\dag}\ket{0}\bra{0}c_{i'}\\
    &\hspace{1em}{-\text{i}}\sum_{i\in A}\sum_{i'\in B}\kuohao{U\mathcal{U}}_{\gamma i'}\kuohao{\mathcal{U}^{\dag}U^{\dag}}_{\gamma i}\ket{0}\bra{0}c_{i'}c_{i}\\
    &\hspace{1em}{-\text{i}}\sum_{i\in A}\sum_{i'\in B}\kuohao{U\mathcal{U}}_{i\gamma}\kuohao{\mathcal{U}^{\dag}U^{\dag}}_{i'\gamma}c_{i}^{\dag}c_{i'}^{\dag}\ket{0}\bra{0},
\end{aligned}
\end{equation}
evidently, transforming $\Tilde{\rho}_{\gamma}^{T_{B}^{f}}$ into ${\rho}_{\gamma}^{T_{B}^{f}}$can be achieved through the unitary transformation $c_{i}\to \exp\fkuohao{-2\text{i arg}(\phi_{\gamma i)}}c_{i}$, provided that $i\in B$, where $\phi_{\gamma i}=\kuohao{U\mathcal{U}}_{\gamma i}$. 

Using Eq. (\ref{equivalence after unitary transformation}), the identity $\mathcal{E}\kuohao{\rho_{\gamma}^{T_{B}^{f}}}=\mathcal{E}\kuohao{\Tilde{\rho}_{\gamma}^{T_{B}^{f}}}$ holds, allowing us to compute bipartite negativity in free-fermion systems via the overlap matrix approach. 

In Appendix \ref{app:comparison of random hamiltonian}, we verify the eigenvalues of Eq. (\ref{upt in d+}) through exact diagonalization and obtain consistent results, indicating that a similarity transformation maps $\rho^{T_{B}^{f}}$ to $\Tilde{\rho}^{T_{B}^{f}}$. This confirms that the overlap matrix can be effectively used to analyze the negativity spectrum \cite{negative_spectrum,negativity_hamiltonian}. 

\section{Logarithmic negativity of free fermions system in bipartite case}\label{sec:bipartite entanglement proof and numerical result}

\begin{figure}[t]
    \centering
    \includegraphics[width=\columnwidth]{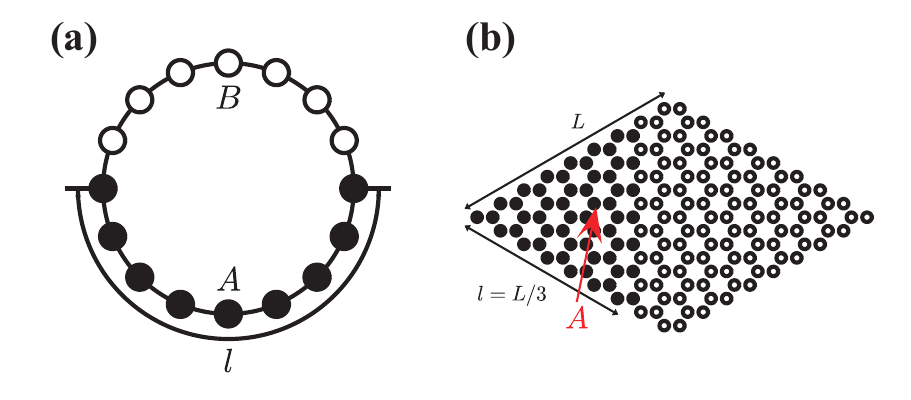}
    \caption{Bipartite system for a chain and a honeycomb lattice. (a) A contiguous region $A$ consisting of $l$ sites is selected within a system of total size $L$. (b) For the honeycomb lattice, the system is partitioned such that one region forms a corner, with each edge consisting of $l=L/3$ unit cells. The entire system contains $L$ unit cells along each lattice vector, allowing us to examine the scaling law in two dimensions.}
    \label{fig:bipartite graph}
\end{figure}

In this section, we consider two free fermion systems. The Hamiltonian we discuss here is 
\begin{equation}\label{trivial free fermions hamiltonian}
    H=-\mu\sum_{i}c_{i}^{\dag}c_{i}-t\sum_{\qiwang{i,j}}\kuohao{c_{i}^{\dag}c_{j}+\text{H.c.}},
\end{equation}
where $\mu>0$ is the chemical potential, $\langle i, j \rangle$ denotes the nearest-neighbor hopping, and $t>0$ is the hopping amplitude. 

\subsection{One-dimensional infinite chain}
Consider an infinite lattice system $\kuohao{L\to\infty}$ with periodic boundary conditions (PBCs), i.e. $c_{N+1}=c_{1}$. In a bipartite setup, as shown in Fig. \ref{fig:bipartite graph}(a), the overlap matrix takes the form of a Toeplitz matrix.
\begin{equation}\label{1DM_A}
    \fkuohao{M_{A}}_{\alpha,\beta}=L^{-1}\sum_{\nu=0}^{l-1}\exp\fkuohao{\text{i}\frac{2\pi}{L}\nu\kuohao{\beta-\alpha}}.
\end{equation}
$\hspace{1.2em}$To obtain analytical results on entanglement, the formula in Eq. (\ref{bipartite log negativity formula}) should be generalized to a continuous form \cite{overlap_for_Sn}. 

\begin{figure}[t]
    \centering
    \includegraphics[width=\linewidth]{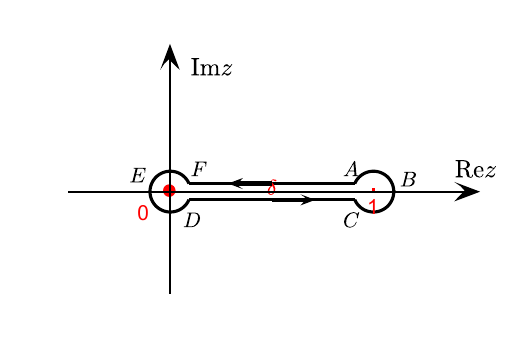}
    \caption{The integral contour in Eq. (\ref{entanglement quantity formula1}) consists of arcs $\overrightarrow{FED}$ and $\overrightarrow{CBA}$ with radii $\epsilon\to 0$. The midpoint of these arcs originate from the range of eigenvalues of the overlap matrix. The imaginary part of $\overrightarrow{AF}$($\overrightarrow{DC}$) is given by $\delta=0^{+}\kuohao{-\delta=0^{-}}$. }
    \label{fig:integral contour}
\end{figure}

$\hspace{1.2em}$If we define the relation of the eigenvalues $P_{\gamma}$ and its contribution to logarithmic negativity as $f\kuohao{P_{\gamma}}$, i.e., $f\kuohao{P_{\gamma}}=\ln\fkuohao{1+2\sqrt{P_{\gamma}\kuohao{1-P_{\gamma}}}}$, then the logarithmic negativity formula becomes
\begin{equation}
\begin{aligned}
    \mathcal{E}&=\sum_{\gamma=1}^{M}f(P_{\gamma})\\
    &=\oint\frac{\text{d}\lambda}{2\pi\text{i}}\sum_{\gamma=1}^{M}\frac{f(\lambda)}{\lambda-P_{\gamma}}\\
    &=\oint\frac{\text{d}\lambda}{2\pi\text{i}}f(\lambda)\frac{\text{d}\ln\ D_{A}(\lambda)}{\text{d}\lambda}, 
\end{aligned}
\end{equation}
where the integral contour is shown in Fig. \ref{fig:integral contour}, and $D_{A}\kuohao{\lambda}=\det\kuohao{\Tilde{M}_{A}=\lambda\mathds{1}-M_{A}}$ represents the characteristic polynomial of the overlap matrix, which is also a Toeplitz matrix. The Fisher-Hartwig conjecture \cite{Fisher-Hartwig_original,detailed_cal-with_Fisher-Hartwig1,detailed_cal-with_Fisher-Hartwig2,colloquium_of_area_law,example_of_Fourier_component_for_Toeplitz_matrix,Fisher-Hartwig_supp1} facilitates the evaluation of the determinant of the Toeplitz matrix $\Tilde{M}_{A}$ revealing that entanglement follows a volume law in a one-dimensional infinite chain (see Appendix \ref{app_b} for a detailed derivation)
\begin{equation}\label{E_N analytical}
\begin{aligned}
    \mathcal{E}&\approx\frac{1}{\pi^{2}}\int_{0}^{1}\text{d}x\frac{f(x)}{x(1-x)}\left[ \ln\ L+\ln\kuohao{2\abs{\sin \frac{1}{2}k_{F}}} \right]\\
    &=\frac{1}{2}\left[ \ln\ L+\ln\kuohao{2\abs{\sin \frac{1}{2}k_{F}}} \right],
\end{aligned}
\end{equation}
where $k_{F}=\frac{l}{L}2\pi$. When $l\ll L$, the logarithmic negativity follows $\mathcal{E}\sim \frac{1}{2}\ln l+C$, where $C$ is a constant, consistent with Ref. \cite{finite_temp_EN,boson_EN_CFT} for central charge $c=1$. This is also supported by the numerical results in Fig. \ref{fig:scaling law}(a). For higher-dimensional systems, such as a hypercube lattice $\fkuohao{0,L_{1}}\times\fkuohao{0,L_{2}}\times\cdots\times\fkuohao{0,L_{d}}$, the Widom conjecture\cite{Widom_conjecture} predicts that the entanglement scales as $\sim L^{d-1}\ln L$ when region $A$ is chosen as $\fkuohao{0,z_{1}L}\times\fkuohao{0,z_{2} L}\times\cdots\times\fkuohao{0,z_{d}L}$ with $z_{i}<1$ for $i=1,...,d$. This result also aligns with previous studies\cite{higher_dimension_1,detailed_cal-with_Fisher-Hartwig2,higher_dimension_2}.

\subsection{Honeycomb lattice}
\begin{figure*}[t]
    \centering
    \includegraphics[width=\columnwidth]{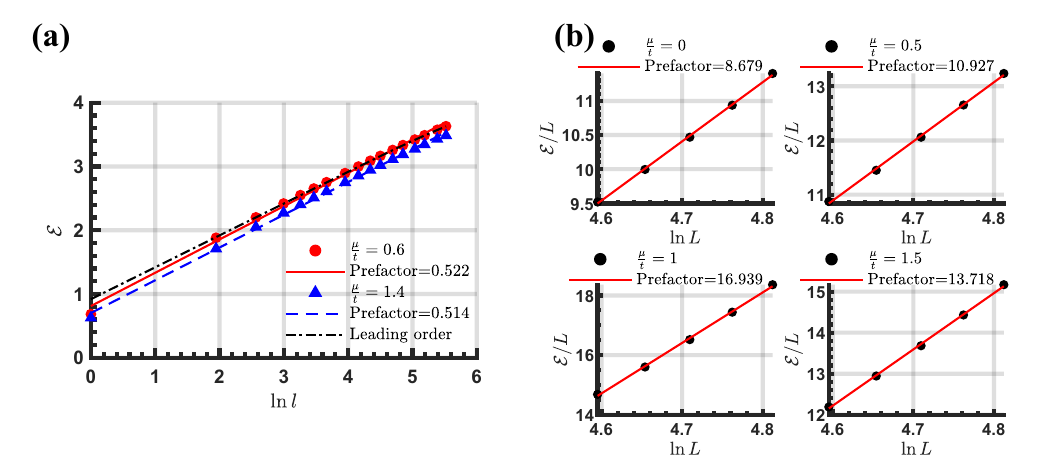}
    \caption{Logarithmic negativity $\mathcal{E}$ in different lattice systems. (a) $\mathcal{E}$ as a function of $\ln l$ in a one-dimensional infinite chain for a bipartite system. The total chain length is set to $1000$ sites. Two parameter choices are compared with the analytical formula (black line), both confirming the linear relationship between $\mathcal{E}$ and the logarithm of the subsystem size.  (b) $\mathcal{E}/L$ vs $\ln L$ in a honeycomb lattice. The results indicate that logarithmic negativity follows the two-dimensional scaling law $\mathcal{E}\sim L\ln L$. The slope varies for different $\mu/t$, reflecting changes in the Fermi surface \cite{data_available}. }
    \label{fig:scaling law}
\end{figure*}

From Fig. \ref{fig:scaling law}(b), the logarithmic negativity of the honeycomb lattice follows the scaling law $\sim L\ln L$, suggesting that the Widom conjecture remains valid. The scaling factor varies depending on the shape and structure of the Fermi surface. This demonstrates that both examples are consistent with the Gioev-Klich-Widom scaling law \cite{higher_dimension_1,Widom_conjecture,Yepeng2024nc}. 

\section{Logarithmic negativity of free fermions system in tripartite case}\label{sec:tripartite entanglement}
Like in Ref. \cite{overlap_for_EN}, when the three overlap matrices $M_{A_{1}},\ M_{A_{2}}$, and $M_{B}$ are simultaneously diagonalizable, their eigenvalues, denoted as $\hkuohao{P_{A_{1}\gamma}},\hkuohao{P_{A_{2}\gamma}},\hkuohao{P_{B\gamma}}$, respectively, determine the logarithmic negativity
\begin{widetext}
\begin{equation}\label{tripartite log negativity formula 1}
\begin{aligned}
    \mathcal{E}&=\sum_{\gamma}\ln\left[ P_{A_{1}\gamma}+P_{A_{2}\gamma}+\sqrt{\frac{1}{2}\kuohao{P_{B\gamma}^{2}+2P_{A_{1}\gamma}P_{A_{2}\gamma}+P_{B\gamma}\sqrt{P_{B\gamma}^{2}+4P_{A_{1}\gamma}P_{A_{2}\gamma}}}}\right.\\
    &\hspace{4em}\left. +\sqrt{\frac{1}{2}\kuohao{P_{B\gamma}^{2}+2P_{A_{1}\gamma}P_{A_{2}\gamma}-P_{B\gamma}\sqrt{P_{B\gamma}^{2}+4P_{A_{1}\gamma}P_{A_{2}\gamma}}}} \right],
\end{aligned}
\end{equation}
\end{widetext}
where $P_{B\gamma}$ represents the probability of the $\gamma$-th particle in region $B$, and fulfills the identity $P_{A_{1}\gamma}+P_{A_{2}\gamma}+P_{B\gamma}=1$. A key distinction of this formula compared with that in Ref. \cite{overlap_for_EN} arises from the nature of the phase factor in the fermionic partial transpose. Significantly, it simplifies to Eq. (\ref{bipartite log negativity formula}) when region $B$ contains no lattice sites, leading to $P_{B\gamma}=0$ for all $\gamma$. 

\begin{figure}[t]
    \centering
    \includegraphics[width=\columnwidth]{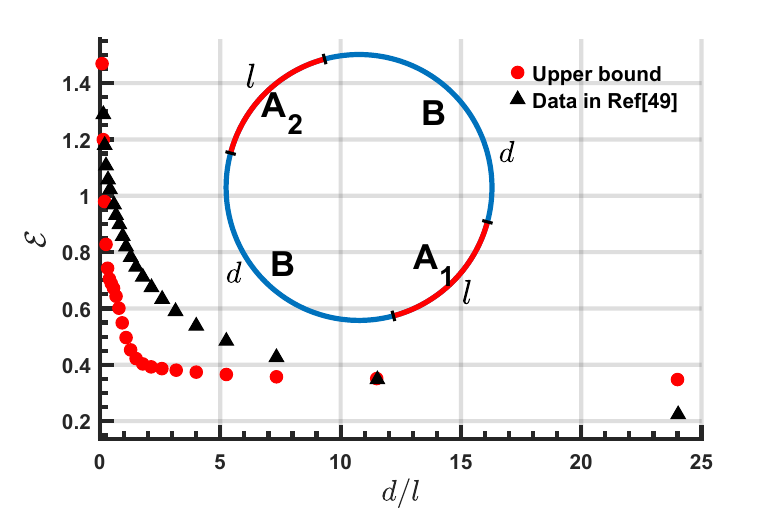}
    \caption{Comparison of the logarithmic negativity in Ref. \cite{overlap_for_EN} with its upper bound in the $M=5$ case, highlighting the necessity of a careful approach in the partial transpose procedure. We consider a one-dimensional homogeneous chain of length $L=100$ with periodic boundary conditions. The specific tripartite geometry used in our analysis is illustrated in the inset figure \cite{data_available}.}
    \label{fig:error in overlap}
\end{figure}

When three overlap matrices cannot be simultaneously diagonalized, previous studies \cite{overlap_for_EN} faced an exponential complexity issue in calculating logarithmic negativity. The Green's function approach provides an upper bound for the logarithmic negativity of free-fermion systems in the bPT framework. Using this method, we find that although the numerical results in Ref. \cite{overlap_for_EN} capture the entanglement scaling law, they violate this upper bound. The upper bound of the mixed-state logarithmic negativity is \cite{Eisler2015,Eisert2018prb,Shapourian2017prb,finite_temp_EN}
\begin{equation}\label{upper bound formula}
    \mathcal{E}\kuohao{\rho_{A}^{T_{A_{1}}^{b}}}\leq\mathcal{E}\kuohao{\rho_{A}^{T_{A_{1}}^{f}}}+\ln\sqrt{2},
\end{equation}
where $\rho_{A}^{T_{A_{1}}^{b}}$ represents the reduced density matrix after bPT. In Fig. \ref{fig:error in overlap}, we demonstrate that the logarithmic negativity reported in Ref. \cite{overlap_for_EN} exceeds the upper bound for the $M=5$ case in the one-dimensional homogeneous chain. 

Our clarification is as follows: While $\Tilde{\rho}^{T_{B}^{f}}$ in the overlap matrix approach connects to $\rho^{T_{B}^{f}}$ in Fock space via a similarity transformation for pure states, this does not hold for mixed states, which are classical mixtures of pure states. The transformations of different pure states vary, meaning that their classical mixtures disrupt the similarity transformation between $\Tilde{\rho}^{T_{B}^{f}}$ and $\rho^{T_{B}^{f}}$. As a result, the findings in Ref. \cite{overlap_for_EN} are not totally exact.

\section{Conclusions}\label{sec:conclusion}
In this study, we have presented a detailed analysis of the overlap matrix approach for computing the pure-state entanglement in free-fermion systems, with a particular emphasis on the partial trace and fermionic partial transpose operations within the overlap matrix framework. In the bipartite case, we demonstrated that for pure states, the eigenvalues of $\rho_A$ remain invariant under the overlap matrix approach. This property establishes the overlap matrix approach as a semiglobal mapping capable of preserving the eigenvalues of $\rho_A$, providing a significant advantage in entanglement computations. Moreover, we derived an analytical formula for the bipartite logarithmic negativity, offering a concise and computationally efficient method for quantifying entanglement in such systems.

\begin{figure}[b]
    \centering
    \includegraphics[width=\linewidth]{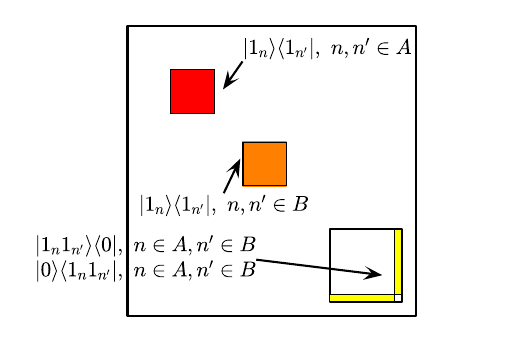}
    \caption{The block diagonal structure of $\Tilde{\rho}_{i}^{T_{B}^{f}}$ and $\rho_{i}^{T_{B}^{f}}$. }
    \label{fig:block diag structure of rho}
\end{figure}

For mixed states, however, the situation is more complex. Using tripartite geometries as a representative example, we identified a critical limitation in the overlap matrix mapping: the logarithmic negativity reported in Ref. \cite{overlap_for_EN} exceeds the theoretical upper bound. This discrepancy arises because the partial transpose operation in the overlap matrix mapping preserves eigenvalues only for pure states. For mixed states, which are classical mixtures of pure states, this equivalence breaks down due to the inconsistent  transformations of individual pure states under the overlap matrix approach. This finding underscores the need for careful consideration when applying overlap matrix approach in the context of mixed states.

In summary, in this work, we highlight both the strengths and limitations of the overlap matrix approach. For bipartite entanglement, this method proves to be a more effective and reliable strategy than the Green's function approach, particularly due to its ability to preserve eigenvalues under semilocal mappings. However, the framework for tripartite entanglement remains incomplete, as the overlap matrix approach encounters challenges when applied to mixed states. In this context, the Green's function approach remains the preferred choice for numerical studies due to its broader applicability.

Beyond these specific findings, in this study, we introduce an additional perspective on entanglement research. We have demonstrated that the partial trace operation in the overlap matrix approach preserves the eigenvalues of the reduced density matrix, a property that has important implications for future research. This insight provides a foundation for further refinement and application of the overlap matrix approach, potentially advancing the field of entanglement computation in free-fermion systems.

\section*{Acknowledgments}
We thank Fo-Hong Wang for helpful discussion. This work was supported by the National Natural Science Foundation of China (Grants No. 12447103 and No. 12274289), the National Key R\&D Program of China (Grants No. 2022YFA1402702 and No. 2021YFA1401400), the Innovation Program for Quantum Science and Technology (under Grant No. 2021ZD0301902), the Yangyang Development Fund, Shanghai Jiao Tong University 2030 Initiative, and startup funds from SJTU. The computations in this paper were run on the Siyuan-1 and $\pi$ 2.0 clusters supported by the Center for High Performance Computing at Shanghai Jiao Tong University.

\section*{data availability}
The data that support the findings of this article are openly available \cite{data_available}, embargo periods may apply.

\onecolumngrid
\appendix
\section{Bipartiteoverlapmatrixapproach}\label{app_a}
\subsection{The range of eigenvalues of overlap matrix. }Here, we will prove that the eigenvalues of the overlap matrix lie within the range $\fkuohao{0,1}$. To demonstrate this, it is convenient to adopt the wave function representation. After a unitary diagonalization, the overlap matrix takes the form
\begin{equation}
\begin{aligned}
    \fkuohao{\mathcal{U}M_{A}\mathcal{U}^{\dag}}_{\alpha\beta}&=\sum_{k,k'}\mathcal{U}_{\alpha k}\fkuohao{M_{A}}_{kk'}\fkuohao{\mathcal{U}^{\dag}}_{k'\beta}\\
    &=\sum_{kk'}\int_{\mathbf{r}\in A}\mathcal{U}_{\alpha k}\phi^{*}_{k}\kuohao{\mathbf{r}}\phi_{k'}\kuohao{\mathbf{r}}\fkuohao{\mathcal{U}^{\dag}}_{k'\beta}\text{d}\mathbf{r}.
\end{aligned}
\end{equation}
A new set of eigenstates $\psi_{\alpha}=\sum_{k}\fkuohao{\mathcal{U}^{\dag}}_{k\alpha}\phi_{k}\kuohao{\mathbf{r}}$ is constructed, preserving both orthogonality and normalization. Consequently, the integral of the squared modulus $\psi_{\alpha}$ over region $A$ cannot exceed $1$.
    
\subsection{The eigenvalues of $\Tilde{\rho}_{i}^{T_{B}^{f}}$. }Here we will prove that the eigenvalues of $\Tilde{\rho}_{i}^{T_{B}^{f}}$ in Eq. (\ref{rho_i^T in overlap matrix mapping}) are identical to those of $\rho_{i}^{T_{B}^{f}}$ in Eq. (\ref{rho_i^T in Fock space}). By rearranging the basis order, both matrices can be block diagonalized into the form shown in Fig. \ref{fig:block diag structure of rho}. The first two blocks, which contain nonzero elements (i.e., the red and orange blocks), are completely identical in both density matrices. Therefore, we only need to analyze the characteristic polynomial of the yellow block. 

$\hspace{1.2em}$The characteristic polynomial has a general form
\begin{small}
\begin{equation}
\begin{aligned}
    \det\kuohao{\begin{matrix}
        -\lambda   &          &        &         & b_{1}\\
                   & -\lambda &        &         & b_{2}\\
                   &          & \ddots &         & \vdots\\
                   &          &        & \ddots  & b_{N-1}\\
        a_{1}      & a_{2}    & \cdots & a_{N-1} & -\lambda
    \end{matrix}}&=-\lambda\det\kuohao{\begin{matrix}
        -\lambda &        &         & b_{2}\\
                 & \ddots &         & \vdots\\
                 &        & \ddots  & b_{N-1}\\
        a_{2}    & \cdots & a_{N-1} & -\lambda
    \end{matrix}}+\kuohao{-1}^{N}a_{1}\det\kuohao{\begin{matrix}
                  &        &           & b_{1}\\
         -\lambda &        &           & b_{2}\\
                  & \ddots &           & \vdots\\
                  &        & -\lambda  & b_{N-1}\\
    \end{matrix}}\\
    &=-\lambda\det\kuohao{\begin{matrix}
        -\lambda &        &         & b_{2}\\
                 & \ddots &         & \vdots\\
                 &        & \ddots  & b_{N-1}\\
        a_{2}    & \cdots & a_{N-1} & -\lambda
    \end{matrix}}+\kuohao{-1}^{2N-1}a_{1}b_{1}\det\kuohao{\begin{matrix}
         -\lambda &        &         \\
                  & \ddots &         \\
                  &        & -\lambda\\
    \end{matrix}}\\
    &=\cdots\\
    &=\kuohao{-1}^{2N-1}\kuohao{-\lambda}^{N-2}a_{1}b_{1}+\kuohao{-1}^{2N-3}\kuohao{-\lambda}^{N-2}a_{2}b_{2}+\cdots+\kuohao{-1}^{2\cdot 2-1}\kuohao{-\lambda}^{N-2}a_{N-1}b_{N-1}\\
    &\equiv f\kuohao{a_{1}b_{1};a_{2}b_{2},\cdots,a_{N-1}b_{N-1}}.
\end{aligned}
\end{equation}
\end{small}
$\hspace{1.2em}$This demonstrates that the characteristic polynomial depends on the sequence $\hkuohao{a_{1}b_{1},a_{2}b_{2},...,a_{N-1}b_{N-1}}$. Evidently, the values of $a_{i}b_{i}$ in $\rho_{i}^{T_{B}^{f}}$ are identical to those in $\Tilde{\rho}_{i}^{T_{B}^{f}}$. Consequently, the characteristic polynomial and the eigenvalues remain unchanged.

\section{Fisher-Hartwig conjecture}\label{app_b}
Following Ref. \cite{detailed_cal-with_Fisher-Hartwig1}, we now proceed with the proof of Eq. (\ref{E_N analytical}).

In Eq. (\ref{1DM_A}), the overlap matrix is a Toeplitz matrix which indicates that the matrix elements $\fkuohao{M_{A}}_{i,j}$ depend only on the value of $i-j$, i.e., $\fkuohao{M_{A}}_{i,j}=f\kuohao{i-j}$, where the function $f$ depends on the specific context. The Toeplitz matrix, denoted as $T_{L}\fkuohao{\phi}$ in general, is generated by a function $\phi\kuohao{\theta}$ if its entries are given by the Fourier coefficients of $\phi\kuohao{\theta}$
\begin{equation}
    T_{L}\fkuohao{\phi}=\frac{1}{2\pi}\int_0^{2\pi}\phi(\theta)\exp\fkuohao{-\mathrm{i}\kuohao{i-j}\theta}\mathrm{d}\theta,\quad i,j=1,\ldots,L-1.
\end{equation}
In our case the overlap matrix
\begin{equation}
\begin{aligned}
    \phi_{l}&\approx \int_{-\frac{1}{2}+\frac{l}{L}}^{\frac{1}{2}}\exp\kuohao{-\text{i}2\pi l\nu}\text{d}\nu\\
    &=\frac{1}{2\pi}\int_{-\pi+\frac{l}{L}\cdot 2\pi}^{\pi}\exp\kuohao{-\text{i}l\nu}\text{d}\nu\\
    &=\frac{1}{2\pi}\text{e}^{\text{i}l\pi}\int_{0}^{2\pi}\phi(\theta)\exp\kuohao{-\text{i}l\theta}\text{d}\theta,
\end{aligned}
\end{equation}
where
\begin{equation}
    \phi\kuohao{\theta}=\begin{cases}
        0 & \text{if }\theta\in\fkuohao{0,k_{F}}\\
        1 & \text{if }\theta\in\fkuohao{k_{F},2\pi}
    \end{cases},\ k_{F}=\frac{l}{L}2\pi
\end{equation}
For the Toeplitz matrix $\Tilde{M}_{A}=\lambda\mathds{1}-M_{A}$, the generating function is obtained by replacing $0$ with $\lambda$ and $1$ with $\lambda-1$. Regarding the singularities of the generating function, it can be decomposed into a product of functions with known asymptotic behavior, which facilitates the application of the Fisher-Hartwig conjecture
\begin{equation}
    \phi(\theta)=\psi(\theta)\prod_{r=1}^{R}t_{\beta_{r},\theta_{r}}(\theta)u_{\alpha_{r},\theta_{r}}(\theta),
\end{equation}
where
\begin{equation}
\begin{aligned}
    t_{\beta_{r},\theta_{r}}(\theta)&=\exp[-i\beta_{r}(\pi-\theta+\theta_{r})],\quad\theta_{r}<\theta<2\pi+\theta_{r}\\
    u_{\alpha_{r},\theta_{r}}(\theta)&=[2-2\cos(\theta-\theta_{r})]^{\alpha_{r}}.\quad\Re\alpha_{r}>-\frac{1}{2}
\end{aligned}
\end{equation}
In our case, the jump discontinuity of the generating function is characterized by $R=2$, with parameters $\beta=-\beta_{1}=\beta_{2}=\frac{1}{2\pi\text{i}}\ln\ \frac{\lambda-1}{\lambda}$ and the singular points located at $\theta_{1}=k_{F}$ and $\theta_{2}=0$. The smooth part of the generating function is given by $\psi\kuohao{\theta}=(\lambda-1)\left( \frac{\lambda-1}{\lambda} \right)^{-k_{F}/2\pi}$.

Then the Fisher-Hartwig conjecture illustrates that the determinant of a Toeplitz matrix
\begin{equation}\label{Fisher-Hartwig conjecture}
    D_{L}=\kuohao{\mathscr{F}\fkuohao{\psi}}^{L}\kuohao{\prod_{i=1}^{R} L^{\alpha_{i}^{2}-\beta_{i}^{2}}} \mathscr{E}\fkuohao{\psi,\hkuohao{\alpha_{i}},\hkuohao{\beta_{i}},\hkuohao{\theta_{i}}}\text{ when }L\to\infty,
\end{equation}
\begin{equation}\label{notation1}
    \mathscr{F}[\psi]=\exp\left[\frac1{2\pi}\int_0^{2\pi}\ln\psi(\theta)\mathrm{~d}\theta\right],
\end{equation}
\begin{equation}\label{notation2}
\begin{aligned}
    \mathscr{E}[\psi,\{\alpha_i\},\{\beta_i\},\{\theta_i\}]& =\mathscr{E}[\psi] \prod_{i=1}^R\frac{G(1+\alpha_i+\beta_i)G(1+\alpha_i-\beta_i)}{G(1+2\alpha_i)}\prod_{i=1}^R (\psi_-(\exp(\mathrm{i}\theta_i)))^{-\alpha_i-\beta_i} (\psi_+(\exp(-\mathrm{i}\theta_i)))^{-\alpha_i+\beta_i} \\
    &\hspace{1em}\times\prod_{1\leqslant i\neq j\leqslant R}(1-\exp(\mathrm{i}(\theta_{i}-\theta_{j})))^{-(\alpha_{i}+\beta_{i})(\alpha_{j}-\beta_{j})},
\end{aligned}
\end{equation}
\begin{equation}\label{notation3}
    \mathscr{E}[\psi]=\exp\left[\sum_{k=1}^{\infty}ks_{k}s_{-k}\right],
\end{equation}
\begin{equation}\label{notation4}
    G(1+z)=(2\pi)^{z/2} \exp\fkuohao{-\frac{(z+1)z}{2}-\frac{\gamma_{E}z^{2}}{2}}\prod_{n=1}^{\infty}\hkuohao{\kuohao{1+\frac{z}{n}}^{n} \exp\kuohao{-z+\frac{z^{2}}{2n}}},
\end{equation}
where $s_{k}$ represents the $k$th Fourier coefficient of $\ln\psi\kuohao{\theta}$. In our case since $\psi\kuohao{\theta}$ is a constant independent of $\theta$, we temporarily denote it as $C$, leading to $s_{k}=C\delta\kuohao{k}$. Consequently, the summation simplifies as follows $\sum_{k=1}^{\infty}ks_{k}s_{-k}\rightarrow\int_{0}^{\infty}C^{2}k\delta(k)\delta(-k)\text{d}k=0$. By substituting Eqs. (\ref{notation1})-(\ref{notation4}) into Eq. (\ref{Fisher-Hartwig conjecture}), we arrive at the desired result.
\begin{equation}
\begin{aligned}
    D_{L}(\lambda)&=\left[ (\lambda-1)\left( \frac{\lambda-1}{\lambda} \right)^{-k_{F}/2\pi} \right]^{L}L^{-2\beta^{2}(\lambda)}\exp\fkuohao{-(1+\gamma_{E})\beta^{2}(\lambda)}\prod_{n=1}^{\infty}\left\{ \left[ 1-\frac{\beta^{2}(\lambda)}{n^{2}} \right]^{n}\exp\fkuohao{\frac{\beta^{2}(\lambda)}{n^{2}}} \right\}(2-2\cos k_{F})^{-\beta^{2}(\lambda)},
\end{aligned}
\end{equation}
and 
\begin{equation}
\begin{aligned}
    \frac{\text{d}\ln\ D_{L}(\lambda)}{\text{d}\lambda}&=L\left[ \frac{1}{\lambda-1}-\frac{k_{F}}{2\pi}\left( \frac{1}{\lambda-1}-\frac{1}{\lambda} \right) \right]-\left\{ 2\ln\ L +(1+\gamma_{E})+\sum_{n=1}^{\infty}\left[ \frac{n}{n^{2}-\beta^{2}(\lambda)}-\frac{1}{n} \right]+\ln\ (2-2\cos k_{F})\right\}\frac{\text{d}\beta^{2}(\lambda)}{\text{d}\lambda}\\
    &=L\left[ \frac{1}{\lambda-1}-\frac{k_{F}}{2\pi}\left( \frac{1}{\lambda-1}-\frac{1}{\lambda} \right) \right]-\frac{2\beta(\lambda)}{\pi\text{i}\lambda(\lambda-1)}\left[ \ln\ L+(1+\gamma_{E})+\ln\ (2|\sin \frac{1}{2}k_{F}|)+\Upsilon(\lambda) \right],
\end{aligned}
\end{equation}
where $\Upsilon(\lambda)=\sum_{n=1}^{\infty}\frac{n}{n^{2}-\beta^{2}(\lambda)}-\frac{1}{n}$. For the linear term in $L$, since all of them exist as a first-order pole in $\lambda$, the integral over $\lambda$ ultimately contributes zero to them based on the residue theorem. Therefore, the entanglement quantity is given by
\begin{equation}\label{entanglement quantity formula1}
    \text{Entanglement quantity}=\frac{1}{\pi^{2}}\oint\text{d}\lambda f(\lambda)\frac{\beta(\lambda)}{\lambda(\lambda-1)}\left[ \ln\ L+(1+\gamma_{E})+\ln\ (2|\sin \frac{1}{2}k_{F}|)+\Upsilon(\lambda) \right],
\end{equation}
where the integral contour is shown in Fig. \ref{fig:integral contour}. The residue theorem ensures that the contributions from the integration paths $\overrightarrow{FED}$ and $\overrightarrow{CAB}$ vanish. Consequently, the integral simplifies to
\begin{equation}
\begin{aligned}
    \text{Entanglement quantity}&=\frac{1}{\pi^{2}}\left( \int_{1+\text{i}0^{+}}^{0+\text{i}0^{+}}+\int_{0+\text{i}0^{-}}^{1+\text{i}0^{-}} \right)\text{d}\lambda f(\lambda)\frac{\beta(\lambda)}{\lambda(\lambda-1)}\left[ \ln\ L+(1+\gamma_{E})+\ln\ \kuohao{2\abs{\sin \frac{1}{2}k_{F}}}+\Upsilon(\lambda) \right].
\end{aligned}
\end{equation}
In the complex number field the phase of $\beta\kuohao{\lambda}$ must be considered to ensure a consistent and well-defined formulation
\begin{equation}
\begin{aligned}
    \beta[x(\in\mathbb{R})+\mathrm{i}0^{\pm}]&=\frac{1}{2\pi\mathrm{i}}\left[\mathrm{Ln}\frac{\left(x-1+\mathrm{i}0^{\pm}\right)\left(x-\mathrm{i}0^{\pm}\right)}{x^{2}}\right] \\
    &=-\mathrm{i}W(x)+\left(\frac{1}{2}-0^\pm\right),
\end{aligned}
\end{equation}
where $W\kuohao{x}=\frac{1}{2\pi}\ln\frac{1-x}{x}$. Then the integral
\begin{equation}
\begin{aligned}
    \text{Entanglement quantity}&=\frac{-1}{\pi^{2}}\int_{0}^{1}\text{d}x\frac{f(x)}{x(x-1)}\left[ \ln\ L+(1+\gamma_{E})+\ln\ (2|\sin \frac{1}{2}k_{F}|) \right]\\
    &\hspace{1em}+\frac{1}{\pi^{2}}\sum_{n=1}^{\infty}\int_{0}^{1}\text{d}x\frac{f(x)}{x(x-1)}n\left\{ \frac{\text{i}W(x)-\frac{1}{2}}{n^{2}-[-\text{i}W(x)+\frac{1}{2}]^{2}}+\frac{-\text{i}W(x)-\frac{1}{2}}{n^{2}-[-\frac{1}{2}-\text{i}W(x)]^{2}}+\frac{1}{n^{2}} \right\}\\
    &\equiv E_{1}+E_{2}.
\end{aligned}
\end{equation}
Introducing the $\psi$ function
\begin{equation}
    \psi(x)\equiv\frac{\mathrm{d}}{\mathrm{d}x}\ln\Gamma(x)=-\gamma_{E}+\sum_{n=0}^{\infty}\frac{1}{n+1}-\sum_{n=0}^{\infty}\frac{1}{n+x},
\end{equation}
with the property $\psi(x+1)=\psi(x)+\frac{1}{x}$, the infinite series is simplified as 
\begin{equation}
\begin{aligned}
    E_{2}&=\frac{1}{\pi^{2}}\int_{0}^{1}\text{d}x\frac{f(x)}{x(x-1)}\left( -1 \right)\left[ -\gamma_{E}+\sum_{n=0}^{\infty}\frac{1}{n+1}-\sum_{n=1}^{\infty}\frac{1}{n}-\frac{1}{2}\left\{ \psi\fkuohao{\frac{1}{2}-\text{i}W(x)}+\psi\fkuohao{\frac{1}{2}+\text{i}W(x)} \right\}-1 \right]\\
    &=\frac{-1}{\pi^{2}}\int_{0}^{1}\text{d}x\frac{f(x)}{x(x-1)}\left( -\gamma_{E}-1 \right)+\frac{1}{2\pi^{2}}\int_{0}^{1}\text{d}x\frac{f(x)}{x(x-1)}\left\{ \psi\fkuohao{\frac{1}{2}-\text{i}W(x)}+\psi\fkuohao{\frac{1}{2}+\text{i}W(x)} \right\}.
\end{aligned}
\end{equation}
The first line of $E_{2}$ cancels out with $E_{1}$, leaving the entanglement quantity as
\begin{equation}
\begin{aligned}
    \text{Entanglement quantity}&=\frac{-1}{\pi^{2}}\int_{0}^{1}\text{d}x\frac{f(x)}{x(x-1)}\left[ \ln\ L+\ln\ (2|\sin \frac{1}{2}k_{F}|) \right]\\
    &\hspace{1em}+\frac{1}{2\pi^{2}}\int_{0}^{1}\text{d}x\frac{f(x)}{x(x-1)}\left[ \psi(\frac{1}{2}-\text{i}W(x))+\psi(\frac{1}{2}+\text{i}W(x)) \right].
\end{aligned}
\end{equation}
The first line precisely corresponds to Eq. (\ref{E_N analytical}), while the second line represents the next-to-leading order correction term.

\section{The eigenvalues of the density matrix after partial transpose in the overlap matrix approach}\label{app:comparison of random hamiltonian}
\begin{figure}[H]
    \centering
    \includegraphics[width=\linewidth]{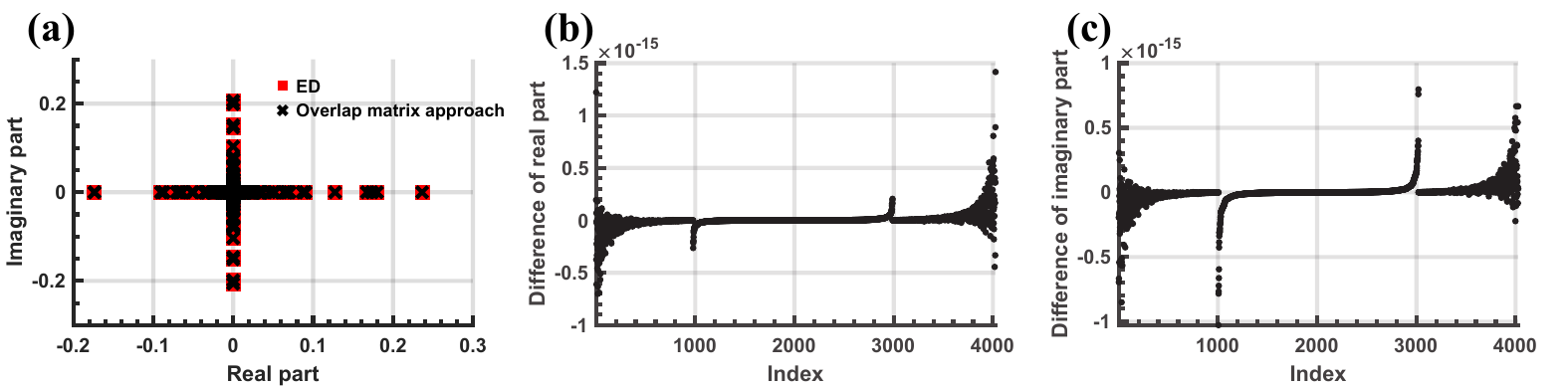}
    \caption{Comparison of the eigenvalues of $\rho^{T_{B}^{f}}$ and $\Tilde{\rho}^{T_{B}^{f}}$: (a) the congruence of eigenvalues between $\rho^{T_{B}^{f}}$ and $\Tilde{\rho}^{T_{B}^{f}}$; (b) difference in real parts of eigenvalues; and (c) difference in imaginary parts of eigenvalues. We select $N=12$ with an even bipartite geometry. The results indicate that the difference is on the order of $10^{-15}$, demonstrating near-perfect agreement \cite{data_available}.}
    \label{fig:comparison of partial transpose}
\end{figure}

In this appendix, we will generalize our result to a random hopping model, i.e.,
\begin{equation}
    H=\sum_{i,j}t_{ij}c_{i}^{\dag}c_{j}=\mathbf{c}^{\dag}t\mathbf{c},
\end{equation}
where the elements of $t$ are random numbers following a uniform distribution while maintaining hermiticity. This can be achieved by generating random matrices $T_{1}$ and $T_{2}$ to construct $T=T_{1}+\text{i}T_{2}$, and then symmetrizing as $t=\frac{T+T^{\dag}}{2}$. Other types of random $t$ can also be generated to further validate our results. Numerical findings confirm that the overlap matrix approach aligns with exact diagonalization, as shown in Fig. \ref{fig:comparison of partial transpose}.

\section{Proof of rationality of partial trace in the overlap matrix approach}\label{app_d}
In this appendix, we expand Eq. (\ref{pt in d+}) into Fock space and compare with the reduced density matrix partially traced in the original basis. We prove that these two equations are completely identical. 

We start from the eigenstate
\begin{equation}
\begin{aligned}
    \ket{\Psi}&=\prod_{\alpha=1}^{M}f_{\alpha}^{\dagger}|0\rangle \\
     & =\prod_{\alpha=1}^{M}\sum_{i=1}^{N}\kuohao{c_{i}^{\dagger}U_{i,\alpha}}\ket{0}\\
     & =\sum_{i_{1},i_{2},\cdots,i_{M}}U_{i_{1},1}U_{i_{2},2}\cdots U_{i_{M},M}c_{i_{1}}^{\dagger}c_{i_{2}}^{\dagger}\cdots c_{i_{M}}^{\dagger}|0\rangle \\
     & =\sum_{1\leq j_{1}<j_{2}<\cdots<j_{M}\leq N}\left[\sum_{\sigma\in S_{M}}\mathrm{sgn}(\sigma)U_{j_{\sigma(1)},1}U_{j_{\sigma(2)},2}\cdots U_{j_{\sigma(M)},M}\right]c_{j_{1}}^{\dagger}c_{j_{2}}^{\dagger}\cdots c_{j_{M}}^{\dagger}|0\rangle \\
     & =\sum_{1\leq j_{1}<j_{2}<\cdots<j_{M}\leq N}\det
    \begin{pmatrix}
    U_{j_{1},1} & U_{j_{1},2} & \cdots & U_{j_{1},M} \\
    U_{j_{2},1} & U_{j_{2},2} & \cdots & U_{j_{2},M} \\
    \vdots & \vdots & \ddots & \vdots \\
    U_{j_{M},1} & U_{j_{M},2} & \cdots & U_{j_{M},M}
    \end{pmatrix}c_{j_{1}}^{\dagger}c_{j_{2}}^{\dagger}\cdots c_{j_{M}}^{\dagger}|0\rangle \\
     & \equiv\sum_{J}\det\left(U_{J}\right)|J\rangle,
\end{aligned}
\end{equation}
where in the fourth line $\sum_{\sigma\in S_{M}}$ represents the summation over all elements of the $M$th-order symmetric group $S_{M}$ ($M$ denotes the particle number) and in the last line $J$ implies a permitted sequence of $\hkuohao{j_{1},j_{2},\ldots,j_{M}}$. Accordingly the density matrix
\begin{equation}
\begin{aligned}
    \rho & =\sum_{J,K}\det\left(U_{J}\right)\det\left(U_{K}^{*}\right)|J\rangle\langle K| \\
     & =\sum_{J,K}\det\left(U_{J}\right)\det\left(U_{K}^{*}\right)|J_{A},J_{B}\rangle\langle K_{A},K_{B}|.
\end{aligned}
\end{equation}
The partial trace operation ensures that the particles in region $B$ remain the same in both bra and ket states, i.e. $J_{B}=K_{B}$. Then we decompose the numerous terms in $\rho_{A}$ based on the particles in region $A$
\begin{equation}
    \rho_{A}=\sum_{m=0}^{\min(M,|A|)}\sum_{\substack{J_{A},K_{A}\subseteq A\\|J_{A}|=|K_{A}|=m}}\left[\sum_{\substack{J_{B}\subseteq B\\|J_{B}|=M-m}}\det(U_{J_{A}\cup J_{B}})\det(U_{K_{A}\cup J_{B}})^{*}\right]|J_{A}\rangle\langle K_{A}|.
\end{equation}
It should be noted that the maximum particle number in region $A$ is $\min\kuohao{M,\abs{A}}$, where $\abs{A}$ represents the the number of sites in region $A$. From Sec. \ref{sec:bipartite entanglement review}, we also know that rotating $\hkuohao{f_{\alpha}}_{\alpha=1}^{M}$ to $\hkuohao{d_{Ai}}_{i=1}^{M}$ and $\hkuohao{d_{Bi}}_{i=1}^{M}$ does not alter the density matrix. Hence, the reduced density matrix can also be expressed as 
\begin{equation}
    \rho_{A}=\sum_{m=0}^{\min(M,|A|)}\sum_{\substack{J_A,K_A\subseteq A \\|J_A|=|K_A|=m}}\left\{\sum_{\substack{J_B\subseteq B \\|J_B|=M-m}}\det\fkuohao{(U\mathcal{U})_{J_A\cup J_B}}\det\fkuohao{(U\mathcal{U})_{K_A\cup J_B}}^*\right\}|J_A\rangle\langle K_A|.
\end{equation}
Then we expand Eq. (\ref{pt in d+}):
\begin{equation}\label{rho_A in overlap matrix mapping}
\begin{aligned}
    \tilde{\rho}_{A} & =\mathrm{Tr}_{B}\tilde{\rho} \\
     & =\prod_{\gamma=1}^{M}\left(P_{\gamma}|1_{A_{\gamma}}\rangle\langle1_{A_{\gamma}}|+(1-P_{\gamma})|0_{A_{\gamma}}\rangle\langle0_{A_{\gamma}}|\right) \\
     & =\prod_{\gamma=1}^{M}\left[P_{\gamma}\frac{\sum_{\alpha=1}^{M}\sum_{j\in A}c_{j}^{\dagger}U_{j,\alpha}\mathcal{U}_{\alpha,\gamma}}{\sqrt{P_{\gamma}}}|0\rangle\langle0|\frac{\sum_{\beta=1}^{M}\sum_{j\in A}c_{j}U_{j,\beta}^{*}\mathcal{U}_{\beta,\gamma}^{*}}{\sqrt{P_{\gamma}}}+(1-P_{\gamma})|0\rangle\langle0|\right] \\
     & =\prod_{\gamma=1}^M\left(\sum_{\alpha=1}^M\sum_{j\in A}c_j^\dagger U_{j,\alpha}\mathcal{U}_{\alpha,\gamma}|0\rangle\langle0|\sum_{\beta=1}^M\sum_{k\in A}c_kU_{k,\beta}^*\mathcal{U}_{\beta,\gamma}^*+\sum_{\alpha=1}^M\sum_{j\in B}U_{j,\alpha}\mathcal{U}_{\alpha,\gamma}|0\rangle\langle0|\sum_{\beta=1}^MU_{j,\beta}^*\mathcal{U}_{\beta,\gamma}^*\right),
\end{aligned}
\end{equation}
where the symbol $\tilde{\rho}_{A}$ is introduced to distinguish it from the original reduced density matrix. In the fourth line we express $1-P_{\gamma}$ as $\sum_{j\in B}\sum_{\alpha,\beta=1}^{M}U_{j,\alpha}\mathcal{U}_{\alpha,\gamma}U_{j,\beta}^*\mathcal{U}_{\beta,\gamma}^*$, which can be verified by the identity $\left(\mathcal{U}^\dagger M_B\mathcal{U}\right)_{\alpha\beta}=\delta_{\alpha\beta}\kuohao{1-P_{\alpha}}$. The essential identity
\begin{equation}
\begin{aligned}
    0&=\sum_{\mu\nu}\sum_{i\in B}\mathcal{U}_{\mu\beta}^{*}U_{i\mu}^{*}U_{i\nu}\mathcal{U}_{\nu\alpha} \\
     & =\sum_{i\in B}(U\mathcal{U})_{i\alpha}(U\mathcal{U})_{i\beta}^{*},
\end{aligned}
\end{equation}
helps derive the equation below:
\begin{equation}
    \tilde{\rho}_{A}=\sum_{m=0}^{\min(M,|A|)}\sum_{\substack{J_A,K_A\subseteq A \\|J_A|=|K_A|=m}}\left\{\sum_{\substack{J_B\subseteq B \\|J_B|=M-m}}\det\fkuohao{(U\mathcal{U})_{J_A\cup J_B}}\det\fkuohao{(U\mathcal{U})_{K_A\cup J_B}}^*\right\}|J_A\rangle\langle K_A|,
\end{equation}
from Eq. (\ref{rho_A in overlap matrix mapping}). It is clear to derive it in reverse order. Ultimately, we demonstrate that $\rho_{A}=\Tilde{\rho}_{A}$. 

\section{The upper bound of logarithmic negativity in bPT}
In this appendix, we derive the upper bound formula for logarithmic negativity, as given in Eq. (\ref{upper bound formula}). Based on Ref. \cite{Eisert2018prb}, the upper bound of logarithmic negativity in bPT is
\begin{equation}\label{upper bound in Ref}
    \mathcal{E}\kuohao{\rho^{T_{B}^{b}}}\leq \ln\tilde{\det}\left[\left(\frac{1+\gamma_\times}{2}\right)^{1/2}+\left(\frac{1-\gamma_\times}{2}\right)^{1/2}\right]+\ln\tilde{\mathrm{det}}\frac{1-\gamma^2}{2}+\ln\sqrt{2},
\end{equation}
where the symbol $\Tilde{\det}$ indicates that double degenerate eigenvalues of the corresponding matrix are counted only once. A normalized Gaussian density operator is defined as $\rho_\times=\frac{O_+O_-}{\mathrm{tr}(O_+O_-) }$. For Gaussian operators
\begin{equation}
    \frac{1}{Z_\sigma}\exp\left[\sum_{k,l}\frac{(W_\sigma)_{k,l}m_km_l}{4}\right],
\end{equation}
the covariance matrix
\begin{equation}
    \gamma_\sigma=\tanh\frac{W_\sigma}{2},\quad\exp(W_\sigma)=\frac{1+\gamma_\sigma}{1-\gamma_\sigma}.
\end{equation}
As far as the spectrum of $\rho_{\times}$ is concerned, it can undergo a similarity transformation
\begin{equation}
    \gamma_\times\simeq(\mathds{1}-\gamma_+\gamma_-)^{-1}(\gamma_++\gamma_-).
\end{equation}
For particle-conserved free fermions, we can replace $\gamma$ with the Green's function $G_{i,j}=\qiwang{c_{i}c_{j}^{\dag}}$ and $\gamma_{\sigma}$ with $G_{\sigma}$ and demonstrate that the first two terms of Eq. (\ref{upper bound in Ref}) precisely correspond to the logarithmic negativity in uPT \cite{Eisler2015,Eisert2018prb,Shapourian2017prb,finite_temp_EN}.

\twocolumngrid
\bibliography{main.bib}

\end{document}